\documentclass[twocolumn,superscriptaddress,amsfont,amssymb,amsmath, showpacs,balancelastpage]{revtex4-1} 
\usepackage{graphicx,longtable,mathrsfs,color,array}
\usepackage[hidelinks]{hyperref}
\usepackage[usenames,dvipsnames]{xcolor} 
\usepackage{amssymb,amsmath,mathtools,mathrsfs} 
\usepackage{epsfig,subfigure,placeins,float} 
\usepackage{booktabs,longtable,ctable,multirow} 
\usepackage{exscale,relsize} 
\usepackage[normalem]{ulem} 
\usepackage{enumerate}
\usepackage{times, mathptmx} 

\newcommand{\be}{\begin{equation}}
\newcommand{\ee}{\end{equation}}
\newcommand{\bea}{\begin{eqnarray}}
\newcommand{\eea}{\end{eqnarray}}

\usepackage[stable]{footmisc}

\def\dkmu2{\delta K_{\mu \nu}\delta K^{\mu \nu}}
\def\pmu2{  \phi_{\mu \nu}\phi^{\mu \nu}}

\begin{document}

\title{Quintessence in a quandary: prior dependence in dark energy models}

\author{David J. E. Marsh}
\email{dmarsh@perimeterinstitute.ca}
\affiliation{Perimeter Institute, 31 Caroline St N, Waterloo, ON, N2L 6B9, Canada}
\author{Philip Bull}
\affiliation{Institute of Theoretical Astrophysics, University of Oslo, P.O. Box 1029 Blindern, N-0315 Oslo, Norway}
\author{Pedro G. Ferreira}
\affiliation{Astrophysics, University of Oxford, DWB, Keble Road, Oxford OX1 3RH, UK}
\author{Andrew Pontzen}
\affiliation{Department of Physics and Astronomy, University College London, Gower Street, London WC1E 6BT, UK}
\date{Received \today; published -- 00, 0000}

\begin{abstract}
The archetypal theory of dark energy is quintessence: a minimally coupled scalar field with a canonical kinetic energy and potential. By studying random potentials we show that quintessence imposes a restricted set of priors on the equation of state of dark energy. Focusing on the commonly-used parametrisation, $w(a)\approx w_0+w_a(1-a)$, we show  that there is a natural scale and direction in the $(w_0, w_a)$ plane that distinguishes quintessence as a general framework. We calculate the expected information gain for a given survey and show that, because of the non-trivial prior information, it is a function of more than just the figure of merit. This allows us to make a quantitative case for novel survey strategies. We show that the scale of the prior sets target observational requirements for gaining significant information. This corresponds to a figure of merit FOM$\gtrsim 200$, a requirement that future galaxy redshift surveys will meet.

\end{abstract}


\maketitle



What drives the accelerated expansion of the Universe? Anything with a sufficiently negative equation of state will do. Consequently, there are a vast number of possible models, generically termed `dark energy' (DE). The equation of state can depend on the scale factor, $a$, and is used to parametrise a wide range of these theories. One is left, however, without a clear idea of how accurate observations must be to actually constrain DE.

Consider the commonly-used series expansion of the equation of state, $w\approx w_0+w_a(1-a)$ \cite{Chevallier:2000qy,Linder:2002et}; this is the parameterisation most commonly used by observers. Given finite resources, what is the optimal precision to which we should measure $w_0$ and $w_a$? To tackle this question we need some theoretical input to identify the regions within the $(w_0, w_a)$ plane that would allow us to to have a realistic chance of actually distinguishing physical models of dynamical DE from a cosmological constant

The archetypal physical model of DE is quintessence \cite{wetterich1988, peebles1988, Caldwell:1997ii}: a scalar field with a potential energy that dominates at late times. If one assumes the well-motivated case of a canonical kinetic energy term, different models consist solely of particular choices of potentials. If the scalar field of quintessence is subject to the rules of effective field theory (EFT), for example, the potential is restricted to a particular functional form, with coupling constants of a particular amplitude (modulo the cosmological constant, $\Lambda$, problem). Similar restrictions arise in specific models within particle physics and string theory, such as pseudo-Nambu-Goldstone Bosons (PNGBs) \cite{hill1988,frieman1995} or axions (e.g. \cite{kaloper2009}), moduli of extra dimensional theories (e.g. \cite{marsh2011,cicoli2012b,marsh2012}), and monodromy \cite{silverstein2008,McAllister:2008hb, panda2010}.

\begin{figure}[tb]
\vspace{-0.2em}\includegraphics[width=\columnwidth]{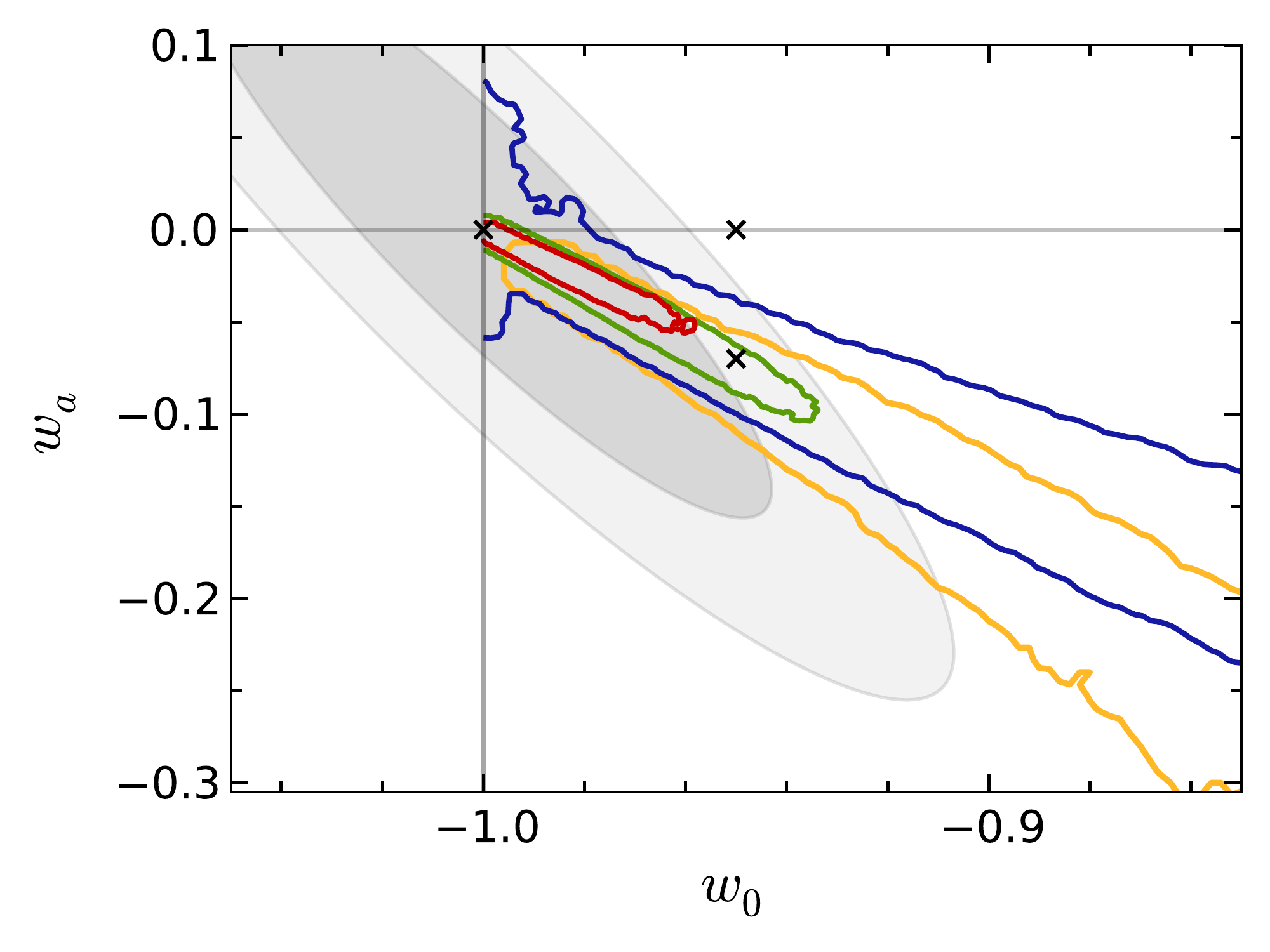}
\vspace{-2.5em}\caption{Quintessence priors in the $(w_0, w_a)$ plane (out to 95\% CL), after loose observational priors have been applied. A remarkably tight structure is observed for all physical models. (Red: EFT; Green: Axion; Blue: Modulus; Yellow: Monomial. All have $\Lambda=0$.) The grey ellipses are predicted constraints for a Dark Energy Task Force Stage IV galaxy redshift survey with a figure of merit of $\sim \!600$. Example fiducial models are shown as crosses.} \vspace{-1.7em}
\label{fig-tracks}
\end{figure}

In this paper we show that quintessence {\it a priori} defines a natural scale and degeneracy direction on the $(w_0, w_a)$ plane when various physical guiding principles are taken into account. This is demonstrated graphically in Fig.~\ref{fig-tracks}, which is a new result of this work. A typical error ellipse for a future galaxy survey with figure of merit (FOM) $\sim 600$ is shown by the filled contours ($1$ and $2\sigma$ regions), and $95\%$ CL regions for the physical quintessence priors are shown by the unfilled contours. The way that these two areas overlap allows us to quantify the information that can actually be gained about DE by undertaking a given survey. 

\emph{Evolution equations ---} The evolution equations are
\bea
3 \left(\frac{\dot{a}}{a}\right)^2 &=& \rho_{r,0} a^{-4} \left ( 1 + a /a_\mathrm{eq} \right ) + \frac{1}{2}\dot{\phi}^2 + A \mathcal{P}(\phi) \, ,\nonumber\\
-6 \left ( \frac{\ddot{a}}{a}\right ) &=& \rho_{r,0} a^{-4} \left ( 2 + a /a_\mathrm{eq}\right ) + 2 \left [ \dot{\phi}^2 - A \mathcal{P}(\phi) \right ] \, ,\nonumber\\
\ddot{\phi} &=& - 3 \dot{\phi}\dot{a}/a - A \mathcal{P}_{,\phi}, \nonumber
\eea
where $\mathcal{P}(\phi)$ is the dimensionless functional form of the quintessence potential and $A$ is its overall scale, $V(\phi) = A M_P^2 M_H^2 \mathcal{P}(\phi)$. We work in units of the reduced Planck mass (energy scale), $M_P = 1 / \sqrt{8 \pi G} = 2.435 \times 10^{27} \mathrm{eV}$, and the Hubble rate (time scale),
$M_H = 100 ~\mathrm{km}\,\mathrm{s}^{-1}\,\mathrm{Mpc}^{-1} = 2.13 \times 10^{-33} \mathrm{eV}$. We have used the redshift of matter-radiation equality, $z_\mathrm{eq}=1/a_\mathrm{eq}-1$ (which we later marginalise over), and the CMB temperature today ($T_\mathrm{CMB} = 2.725$ K) to fix the relative matter and radiation densities,
$\rho_{r,0} = 1.681 \frac{\pi^2}{15} k_B^4 T^4_\mathrm{CMB} M_P^{-2} M_H^{-2} = {\rho_{M,0}}/{(1 + z_\mathrm{eq})}$, where
the leading numerical factor accounts for photons and three generations of neutrinos with negligible mass. 

The DE equation of state is $w(a) = P_\phi / \rho_\phi\approx w_0 + (1 - a) w_a$. The coefficients can be evaluated directly at $a=1$,
\bea
w_0 =  \frac{\dot{\phi}^2 - 2 A \mathcal{P}(\phi)}{\dot{\phi}^2 + 2 A \mathcal{P}(\phi)}, \
w_a =  \frac{2 A}{\dot{a} \rho^2_\phi} \left [ 3 \mathcal{P}(\phi) \dot{\phi}^2 H + \mathcal{P}_{,\phi} \dot{\phi} \rho_\phi \right ].  \nonumber
\eea
In our units, the fractional density in a given component is $\Omega_X(a) = \rho_X(a) / 3 H^2(a)$. Where relevant, we include the cosmological constant (c.c.) within $V(\phi)$ and hence $w$.

We proceed by Monte Carlo sampling (a) various random functional forms for the potential, (b) the parameters of these functional forms, and (c) the initial conditions of the field. The resulting cosmologies are subjected to loose observational cuts to ensure broad consistency with the real Universe. 

\emph{Functional forms ---} We consider a number of general quintessence potentials with functional forms
\begin{equation}
\mathcal{P}(\phi)=c_\Lambda \xi_\Lambda+f(\phi)+\sum_{n_{\rm min}}^{n_{\rm max}} c_n \xi_n b_n(\phi)\, , \nonumber
\end{equation}
where $c_n$ is a deterministic constant, $\xi_n$ is a random variable, $b_n(\phi)$ is a basis function and $f(\phi)$ is a leading contribution to the potential \footnote{All models here could be considered with $f(\phi)=0$ if $c_n$ are individually specified, but this notation is more compact. Allowing $f(\phi)\neq 0$ and product basis functions can model more general potentials within our scheme, such as monodromy and Albrecht-Skordis/SUGRA \cite{Albrecht:1999rm}.}. The term $c_\Lambda \xi_\Lambda$ takes account of the c.c., with $c_\Lambda=0,1$ switching it off/on. The random coefficients are drawn from a unit Gaussian distribution, $\xi_i\equiv \xi \in \mathcal{N}(0,1)$. All potentials are truncated at finite order $n_{\rm max}$, while $n_{\rm min}$ is model-specific. In this paper we consider various types of potential, summarised in Table~\ref{tab:models}. Free parameters are sampled according to the distributions given in Table~\ref{tab:params}. These distributions are chosen to be sufficiently broad and reasonable to capture a wide range of quintessence behaviours.

{\bf Kac/Weyl} potentials are simple random polynomial functions \cite{terrytao}. These will serve as baseline random potentials, but have no physical motivation.

A {\bf Monomial} potential is an integer power law, with only a leading order part, $f(\phi) = \phi^N$. Although possible physical motivations include possible relation to chaotic inflation \cite{Linde:1983gd}, or as large-field limits of certain monodromy models, our chief reason for including these potentials is simplicity.

{\bf EFT} potentials contain a leading `classical contribution' \footnote{Note that the classical contribution is also suppressed by powers of $\epsilon_F$; this is formally a tuning, and reflects the unknown solution to the `old' c.c. problem \cite{Burgess:2013ara}.}, $f(\phi)=\epsilon_{\rm F}^2\xi_2\phi^2+\epsilon_{\rm F}^4\xi_4\phi^4$, plus a random polynomial of `quantum corrections' expanded in an energy scale parameter, $\epsilon_{\rm F}$. To allow quintessence-like masses and energy densities, one requires $|\phi|>1$, and therefore the EFT must be controlled by a super-Planckian shift symmetry, $F>M_P$ \footnote{This may be problematic to realise in UV models including quantum gravity \cite{arkani-hamed2007}.}. For $\epsilon_{\rm F}=M_P/F<1$, this fixes $c_n=\epsilon_{\rm F}^n$. In order to have the expansion begin at some leading order beyond the classical contribution, $n_{\rm min}=p_E>4$. The number of quantum correction terms is $n_Q=n_{\rm max}-p_E+1$.

The potential for an {\bf Axion/PNGB} is a sum of cosines. We choose $f(\phi)$ such that the leading term contributes no c.c. in the vacuum, as is conventional for axions, and higher-order non-perturbative corrections are suppressed by $\epsilon_\mathrm{NP}<1$. The shift symmetry is controlled by the scale $F>M_P$, so $\epsilon_{\rm F}<1$. 

The potential for a {\bf Modulus} of a higher dimensional theory generically includes exponentials \cite{Freund:1980xh}. There can be leading positive exponentials, with higher-order negative exponentials suppressed by the compactification scale $\epsilon_D=(l M)^{-2}$, where $l\lesssim 10^{-6}{\mathrm  m}$ is a length scale and $M<M_P$ a mass scale, giving $f(\phi)=0$, $b_n(\phi)=\exp (\alpha(p_D-n)\phi)$, $c_n=\epsilon_D^n$ and $n_{\rm min}=0$.

\begin{table}[t]
\begin{center}
{\renewcommand{\arraystretch}{1.5}
\hspace{-0.5em}\begin{tabular}{|c|c|c|c|c|c|c|}
\hline
{\bf Model} &  $b_n(\phi)$ & $c_n$ & $n_{\rm min}$ & $f(\phi)$ & $\phi_i$\\
\hline
Kac & $\phi^n$ & $1$ & $1$ & $0$ & $[-1,1]$\\
Weyl & $\phi^n$ & $1/\sqrt{n!}$ & $1$ & $0$ & $[-1,1]$\\
Mono. & 0 & -- & -- & $ \phi^N$ & $[0,4]$\\
\hline
\parbox[t]{7mm}{\multirow{2}{*}{EFT}} & \parbox[t]{4mm}{\multirow{2}{*}{$\phi^n$}} & \parbox[t]{6mm}{\multirow{2}{*}{$(\epsilon_\mathrm{F})^n$}} & \parbox[t]{4mm}{\multirow{2}{*}{$p_E$}} & $~~\,\,\,\xi_2 \epsilon_{\rm F}^2\phi^2$ & \parbox[t]{15mm}{\multirow{2}{*}{$[-\epsilon_{\rm F}^{-1},\epsilon_{\rm F}^{-1}]$}} \\
 &  &  &  & $+~ \xi_4 \epsilon_{\rm F}^4\phi^4$ & \\
Axion & $\cos(n \epsilon_\mathrm{F} \phi)$ & $(\epsilon_\mathrm{NP})^{n-1}$ & $2$ & $1+\cos \epsilon_{\rm F}\phi$& $~[-\frac{\pi}{\epsilon_{\rm F}},\frac{\pi}{\epsilon_{\rm F}}]~$ \\
~Modulus~ & $e^{\alpha(p_D-n)\phi}$ & $(\epsilon_\mathrm{D})^{n}$ & $0$ & $0$& $[-1,1]$ \\
\hline
\end{tabular} }
\end{center}
\vspace{-1.5em}\caption{Model specifications for the functional form $\mathcal{P}(\phi)$.}
\vspace{-1.0em}
\label{tab:models}
\end{table}

\begin{figure*}[tbh]
\vspace{-1.5em}\includegraphics[width=1.9\columnwidth]{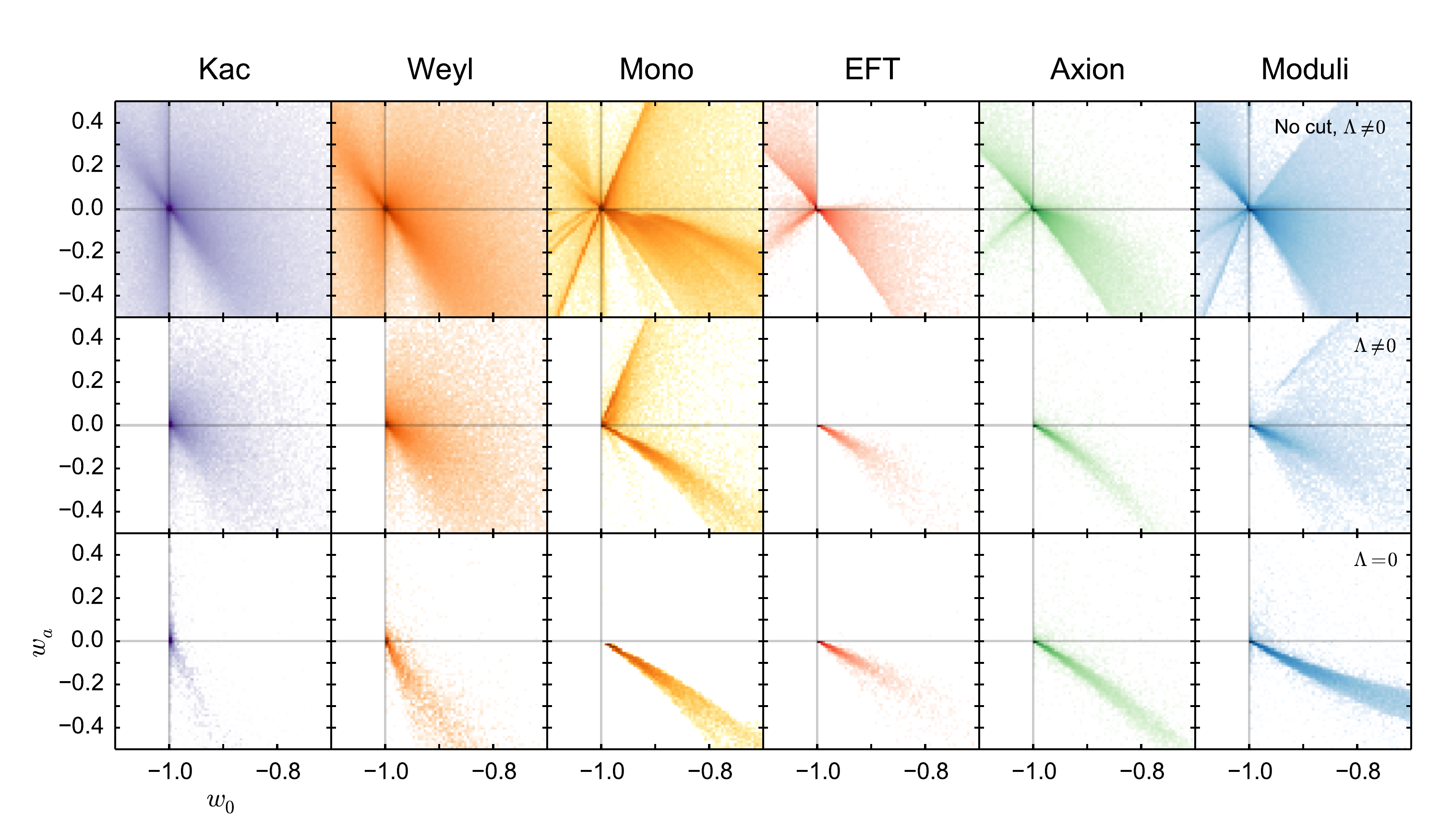}
\vspace{-1.0em}\caption{The (log-scaled) density of prior samples in the $(w_0, w_a)$ plane before (top) and after (middle) the observational cuts, for models with a cosmological constant and $\dot{\phi}_i = 0$. Models with $\Lambda=0$ are shown at the bottom, also after observational cuts. Sharp edges in the cut distributions emerge from the fact that $|w(z)|\leq 1$ when $\Omega_{\rm DE}>0$ drives the current accelerated expansion in quintessence models \cite{Huterer:2006mv}.}\vspace{-1em}
\label{fig-mosaic}
\end{figure*}

\emph{Initial conditions ---} Initial conditions on the field are drawn from a uniform distribution at $a_i = 10^{-2} \,a_\mathrm{eq}$, well before matter-radiation equality. Field displacement can always be reabsorbed in a shift, but total displacement is relevant to the fate of the universe \cite{kallosh2002,kallosh2003} and depends on UV completion \cite{lyth1997}. The natural field range for each of our models is given in Table~\ref{tab:models}. For Kac/Weyl and monomial models we take $\phi\in [-1,1]$ and $\phi_i\in [0,4]$ respectively from demands on energy density, steepness, symmetry and zeros \cite{terrytao}. For EFT controlled by a super-Planckian shift symmetry, the natural range is $[-\epsilon_{\rm F}^{-1},\epsilon_{\rm F}^{-1}]$; for PNGB/axions it is $[-\pi\epsilon_{\rm F}^{-1},\pi\epsilon_{\rm F}^{-1}]$; and for moduli it is $[-1,1]$, emerging from $\epsilon_D<1$ and $1/l<M<M_{P}$ for sub-Planckian compactification. 

\begin{table}[t]
\begin{center}
{\renewcommand{\arraystretch}{1.5}
\begin{tabular}{|c|c|c|c|c|c|c|}
\hline
{\bf Parameter} &  {\bf Model} & {\bf Dist.} \\
\hline
$\log_{10}A$ & All & $\mathrm{U}(-1,1)$ \\
$N$ & Monomial & $\mathrm{U}_{\mathbb{Z}}(1,7)$ \\
$n_{\rm max}$ & ~Kac, Weyl, Ax., Mod.~ & ~$\mathrm{U}_{\mathbb{Z}}(10,20)$~ \\
$n_{\rm Q}$, $p_E$ & EFT & $\mathrm{U}_{\mathbb{Z}}(5,10)$ \\
~$\log_{10}\epsilon_{\rm F,NP,D}$~ & EFT, Ax., Mod.  & $\mathrm{U}(-3,-1)$ \\
$p_D$ & Modulus  & $\mathrm{U}_{\mathbb{Z}}(1,5)$ \\
$\alpha$ & Modulus  & $\mathrm{U}(0,1)$ \\
\hline
\end{tabular} }
\end{center}
\vspace{-1em}
\caption{Parameter distributions for the models in Table~\ref{tab:models}. $\mathrm{U}$ is the uniform distribution, and subscript `$\mathbb{Z}$' indicates that the distribution is over the integers.}
\vspace{-2em}
\label{tab:params}
\end{table}

We observed little difference in the resulting $(w_0, w_a)$ priors based on the prior on $\dot{\phi}\neq 0$ over a large range: Hubble friction damps the field motion at high-$z$. We have also tested our models with log-flat priors on the initial conditions for the field and field velocity, and found that this also had little effect on the $(w_0, w_a)$ priors.
\pagebreak

\emph{Observational cuts ---} 
Models with excessive amounts of early DE are discarded \cite{Pettorino:2013ia} by requiring $\Omega_\mathrm{DE}(z_\mathrm{LSS} \!\approx\! 1090) < 0.042$ \cite{joudaki2012}; we require that the present Hubble rate, $h=H_0/M_H$, is between $0.6<h<0.8$, and the fractional DE density is between $0.6<\Omega_\mathrm{DE}<0.8$. We also put a weak prior on the present-day matter density by sampling $z_\mathrm{eq} \sim \mathrm{U}[2000, 4000]$. We hold $T_\mathrm{CMB}$ fixed and do not vary the neutrino density. We reject any cosmologies that do not reach $a=1$ due to collapse.

These cuts are broader than current observational constraints allow; their purpose is only to act as priors to ensure that we are considering somewhat realistic cosmologies. Typically $1 - 10\%$ of the samples remain after the various cuts are applied, so we draw $\sim\! 10^6$ samples for each model to ensure sufficient statistics. Fig.~\ref{fig-mosaic} shows the Monte Carlo prior samples before and after cuts for each model. 

\emph{Results ---} There is a strong correlation between the equation of state values at different redshifts in quintessence models \cite{Huterer:2006mv}, which is observed as restrictive joint priors on $(w_0, w_a)$ once our broad priors on other cosmological parameters are imposed, as shown in Figs.~\ref{fig-tracks} and \ref{fig-mosaic}. The more typical assumption of independent uniform priors on all of $\{H_0,\Omega_M h^2,w_0,w_a\}$ is not valid for generic physical quintessence models.   

Quintessence models define a narrow strip in the $(w_0,w_a)$ plane, with which certain values (such as the reference point $(-0.95,0)$ \cite{Amendola:2012ys}) are inconsistent. Most of the allowed prior region has $w_a<0$ because acceptable potentials are shallow and typically do not have runaway behaviour as long as the universe is expanding \footnote{The moduli models, having exponential potentials, can more easily support tracking solutions \cite{Ferreira:1997hj} and positive runaways, and so have more prior weight with $w_a>0$.}. In such a potential, sub-dominance of DE at early times and Hubble friction send $w\rightarrow -1$ at high-$z$, while friction lessens at low-$z$, allowing $w$ to become larger. Thus the prior lies near (but not exactly on) the `thawing' region of Ref.~\cite{Caldwell:2005tm} (see also Ref.~\cite{Scherrer:2007pu}). The random quintessence models studied by Ref.~\cite{Huterer:2006mv} constructed using priors in the flow equations were found to be almost entirely `freezing': random evolution constructs arbitrary potentials, distinct from our random physical models (see also Refs.~\cite{Chongchitnan:2007eb, Linder:2007wa}).

Because of this asymptotic behaviour at high-$z$, the $(w_0,w_a)$ parametrisation can fail at $z\gtrsim 2$ for quintessence, apparently predicting $w(z)<-1$. Some of our models lie in this `apparent phantom' region of the $(w_0,w_a)$ plane, but are actually non-phantom for all $z$. The $(w_0, w_a)$ fit should always be taken to have broken down as a description of quintessence above a given $z$ if it predicts $w(z) < -1$. Our priors for this parametrisation are suitable for forthcoming low-$z$ tests of DE; for (e.g.) the CMB, the full scalar field evolution or a different asymptotic fit for $w(z)$ should be used.

To quantify the effect of a non-trivial prior we calculate the information gain over the prior from conducting a given DE survey. The information is a uniquely-motivated quantity for describing the constraining power of a given probability distribution \cite{kullback1951}. The gain in information from conducting a set of observations is sometimes known as the Kullback-Leibler divergence, and for discrete (binned) probability distributions is defined as $\Delta S = \sum_k P_k \log ( P_k / Q_k )$. Here, $k$ labels the bins, $Q_k$ is the prior (i.e. the normalised 2D histogram in $(w_0, w_a)$-space), $P_k = C \mathcal{L}_k Q_k$ is the normalised posterior, and $\mathcal{L}_k$ is the likelihood. In the limit that the likelihood is completely uniform (i.e. uninformative), $\Delta S \to 0$. For the purposes of illustration, we use a Gaussian likelihood with the covariance given by the inverse of a Fisher matrix for a future galaxy redshift survey, $\mathcal{F}$, centred on some fiducial point $(w_0, w_a)|_\mathrm{fid.}$, and marginalised over all other parameters \footnote{For our illustration we approximate the Fisher matrix as being constant as a function of fiducial point, but this would not be the case for a real survey.}. We consider the cases where the fiducial point is fixed and where it is marginalised, defining $\langle \Delta S \rangle = \int Q(x,y) \Delta S(x,y) dx\, dy / \int Q(x, y) dx\, dy$, where $(x, y)$ run over all fiducial values of {$(w_0, w_a)$}. We rescale the covariance matrix by the figure of merit, $\mathrm{FOM} = 1 / \sqrt{\mathrm{det} \,\mathcal{F}^{-1}|_{w_0, w_a}}$, which (loosely) increases with the increasing accuracy of distance measurements from a survey. We also consider the possibility of having an error ellipse that is orthogonal to that of a galaxy survey, which could be achieved in practise by (e.g.) cosmic shear \cite{Linder:2002dt} or redshift drift measurements \cite{Kim:2014uha}. It is also possible to partially rotate the error ellipse of a redshift survey by making an appropriate choice of target redshift and binning.
\begin{figure}[t!]
\includegraphics[width=\columnwidth]{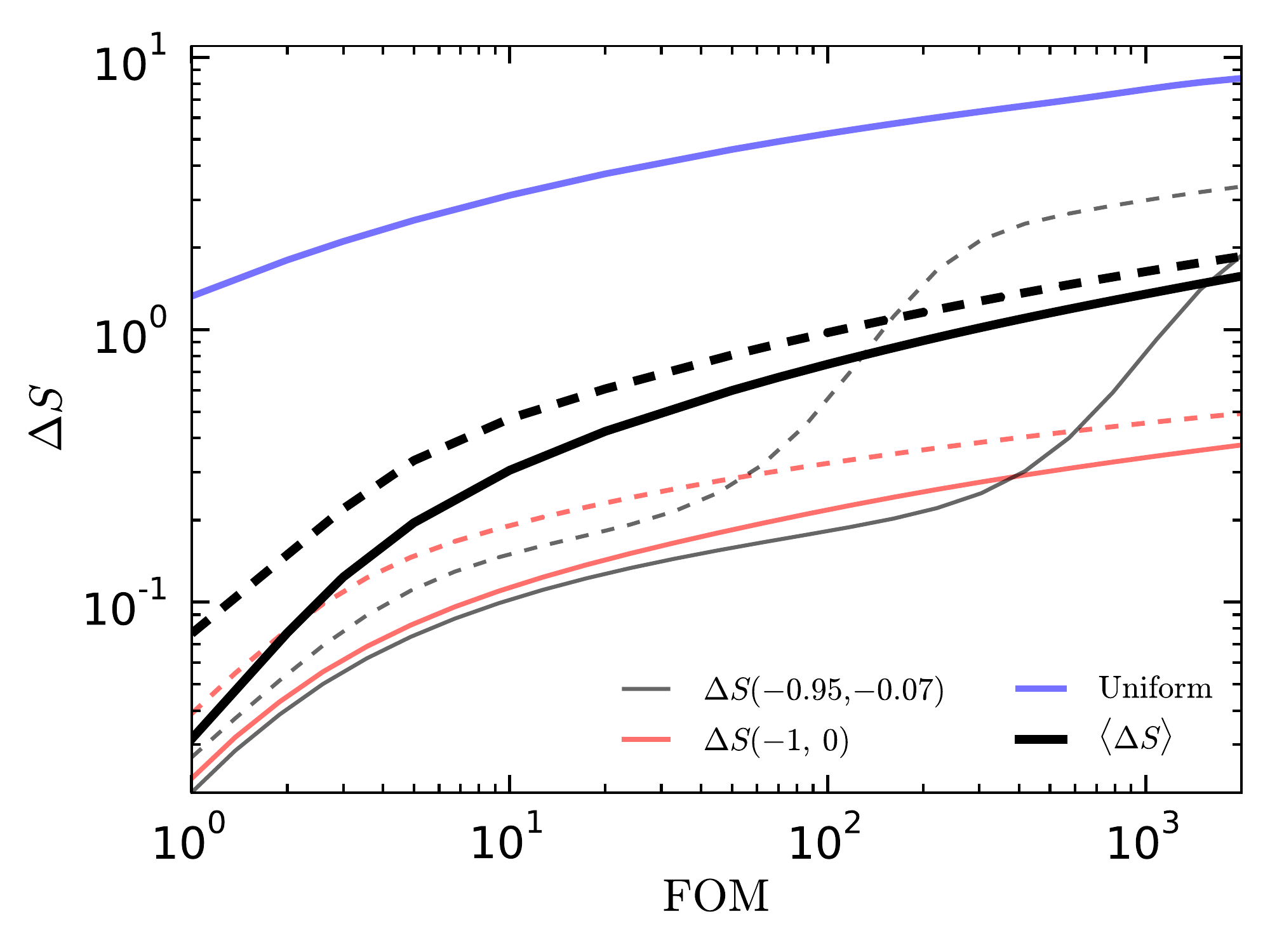}
\vspace{-2em}\caption{Relative entropy as function of figure of merit for a typical future galaxy survey (solid lines), and for the same but with error ellipse rotated by $90^\circ$ (dashed lines). The red and grey lines are for likelihoods fixed at given fiducial values of $(w_0, w_a)$, while the black lines are for $\Delta S$ marginalised over all fiducial values, $\langle \Delta S \rangle$. The thick blue line shows $\Delta S$ for a uniform prior over $(w_0, w_a)$-space, and does not depend on the fiducial point.}
\label{fig-dS}
\vspace{-2em}
\end{figure}

In Fig.~\ref{fig-dS} we show $\Delta S$ as a function of FOM for physical quintessence as a whole (i.e. combining, with equal weights, the normalised prior distributions for all but the unmotivated Kac, Weyl, and Monomial models), and compare this to the $\Delta S$ that would be obtained if uniform priors on $(w_0,w_a)$ were assumed. The value of $\Delta S$ is larger for the uniform prior -- since quintessence disfavours large regions of the $(w_0, w_a)$ plane {\it a priori}, there is less information to be gained from a given survey than if all regions have equal prior probability. With quintessence priors, we also observe features in $\Delta S$ as a function of FOM as the observational error shrinks inside the prior region about a fixed fiducial point (c.f. the results for the point $(-0.95, -0.07)$ in Fig. \ref{fig-dS}). The function marginalised over all fiducial points, $\langle\Delta S\rangle$, does not show such a feature however; this is because the prior is still dominated by the $\Lambda$-like peak at $(w_0,w_a)=(-1,0)$.  

The value $\langle\Delta S\rangle=1$ defines a meaningful scale in FOM to aim for where one begins to gain significant information over the prior (in the example of one dimensional Gaussians with equal mean for the prior and posterior, $\Delta S=1$ corresponds to a posterior with five times smaller $\sigma$ than the prior, and is therefore related to a $5\sigma$ detection threshold). This occurs for FOM~$\approx 200$ for our quintessence priors applied to a galaxy redshift survey. Our reference DETF Stage IV experiment surpasses this requirement. Even if they merely tighten constraints around the c.c. case, observations with this precision are valuable since they can start to rule out significant portions of the prior space of quintessence. 

The orientation of the error ellipse, though unimportant in the uniform prior case, has a substantial effect for the quintessence prior; with the orthogonal ellipse, one always finds a greater information gain. For example, a survey with an orthogonal ellipse and FOM $\sim 100$ offers an equivalent $\langle\Delta S \rangle$ to a standard galaxy redshift survey with a much higher FOM of $\sim 250$. This is due to the near-alignment of the quintessence prior with the typical degeneracy direction of the galaxy survey error ellipse (Fig. \ref{fig-tracks}); an orthogonal ellipse cuts through the prior more effectively.

In this {\it Letter} we considered random, physically-motivated models of quintessence, which were found to impose a specific structure on the DE equation of state. The resulting prior on $(w_0, w_a)$ is only weakly sensitive to the details of how the models are constructed, and is therefore suitable as a guide to regions of observational interest. The value of FOM where $\langle\Delta S\rangle=1$ gives a target for surveys that wish to constrain quintessence. Our results also quantify how surveys at fixed FOM are not equivalent in the amount of information on DE they bring to bear. 

\textit{Acknowledgements ---} We acknowledge C. Burgess, E. Copeland, A. Liddle, J. March-Russell, M.~C.~D. Marsh, P. Marshall and H. Peiris for useful discussions. DJEM acknowledges Oxford University for hospitality. PB is supported by European Research Council grant StG2010-257080, and acknowledges Oxford Astrophysics, Caltech/JPL, and Perimeter Institute for hospitality. PGF acknowledges support from Leverhulme, STFC, BIPAC and the Oxford Martin School. AP acknowledges a Royal Society University Research Fellowship. This research was supported in part by Perimeter Institute for Theoretical Physics. Research at Perimeter Institute is supported by the Government of Canada through Industry Canada and by the Province of Ontario through the Ministry of Economic Development \& Innovation. The computer code used in this paper is publicly available at \url{gitorious.org/random-quintessence}.

\appendix


\bibliographystyle{apsrev4-1}
\bibliography{Quandry}

\end{document}